\documentclass[11pt]{article}
\usepackage{amssymb, latexsym, amsbsy, amsfonts}

\newtheorem{definition}{Definition}[section]
\newtheorem{theorem}{Theorem}[section]

\newtheorem{lemma}[theorem]{Lemma}
\newtheorem{corr}[theorem]{Corollary}
\newtheorem{rem}{Remark}
\newcommand{\remark}{\begin{rem} \normalfont}\def\endremark{\end{rem}}

\newcommand{\cov}{\mathop{\mathrm{cov}}\nolimits}
\newcommand{\supp}{\mathop{\mathrm{supp}}\nolimits}

\newcommand{\vp}{\mathop{\mathrm{{\varphi}}}\nolimits}

\newcommand{\M}{\mathop{\mathrm{{\cal M}}}\nolimits}

\newcommand{\C}{\mathop{\mathrm{{\cal C}}}\nolimits}

\newcommand{\e}{\mathop{\mathrm{{\varepsilon}}}\nolimits}
\newcommand{\p}{\mathop{\mathrm{{\cal P}}}\nolimits}

\newcommand{\Spec}{\mathop{\mathrm{Spec}}\nolimits}

\newenvironment{pf}{\noindent \textit{Proof}: \ignorespaces}{\hfill\ensuremath
{\Box}\par}\newenvironment{pft}{\noindent \textit{Proof of the theorem}: 
\ignorespaces}{\hfill\ensuremath{\Box}\par}

\makeatletter
\@addtoreset{equation}{section}
\makeatother

\newcommand {\R} {\ensuremath{\mathbb{R}}}
\newcommand {\B} {\ensuremath{\mathcal{B}}}

\newcommand {\n}  {\ensuremath {\mathsf{N}}}

\newcommand {\K}  {\ensuremath {\mathcal{K}}}
\newcommand {\h}  {\ensuremath {\mathcal{H}}}
\newcommand {\Z}  {\ensuremath {\mathcal{Z}}}

\newcommand {\s}  {\ensuremath {\mathcal{S}}}

\newcommand {\N}  {\ensuremath {\mathcal{N}}}
\newcommand {\G}  {\ensuremath {\mathcal{G}}}
\newcommand {\g}  {\ensuremath {\mathsf{G}}}
\newcommand {\pp}  {\ensuremath {\mathsf{P}}}

\newcommand {\Ga} {\ensuremath{\Gamma}}

\newcommand {\rh} {\ensuremath{\varrho}}

\begin{document} 
\begin{titlepage}
  \begin{center}
    {\Large \bf The infrared behaviour in Nelson's model of a quantum
      particle coupled to a massless scalar field}
\end{center}
\begin{center}
  
    \small J\'ozsef L\H{o}rinczi$^1$, Robert A. Minlos$^2$ and Herbert Spohn$^1$\\[0.1cm]
\end{center}
\vspace{0.5cm}
    {\it \small $^1$Zentrum Mathematik, Technische Universit\"at M\"unchen} \\
    {\it \small 80290 M\"unchen, Germany} \\
    {\small lorinczi@mathematik.tu-muenchen.de}\\
    {\small spohn@mathematik.tu-muenchen.de}\\ [0.5cm]
    {\it \small $^2$Dobrushin Mathematics Laboratory} \\
    {\it \small Institute for Information Transmission Problems} \\
    {\it \small Bolshoy Karetny per. 19, 101447 Moscow, Russia }\\
    {\small minl@iitp.ru}\\ [1cm]
\vspace{1.5cm}    
\date{}
\begin{abstract} 
We prove that Nelson's massless scalar field model is infrared divergent in three dimensions. In 
particular, the Nelson Hamiltonian has no ground state in Fock space and thus it is not unitarily 
equivalent with the Hamiltonian obtained from Euclidean quantization. In contrast, for dimensions 
higher than three the Nelson Hamiltonian has a unique ground state in Fock space and the two 
Hamiltonians are unitarily equivalent. We also show that the Euclidean Hamiltonian has no spectral 
gap. 
\end{abstract}
\end{titlepage}
\vspace{1cm}
\section{Introduction} 
A quantum particle coupled to a free Bose field is described by the Hamiltonian
\begin{eqnarray}
\label{fs}
H_{\mbox{\tiny{gN}}} 
& = &
\left(-\frac{1}{2} \Delta + V(q)\right) \otimes 1 + 1 \otimes \int_{\R^d} \omega(k) a^*(k) a(k) dk  \\ 
& + & 
\int_{\R^d} \frac{1}{\sqrt{2\omega(k)}} \frac{1}{(2\pi)^{d/2}} \left(\hat \rh(k) e^{ik \cdot q} \otimes 
a(k) + \hat\rh(-k) e^{-ik\cdot q} \otimes a^*(k) \right) dk.
\nonumber
\end{eqnarray}
In the first term we recognize the Schr\"odinger operator $H_{\mathrm{\tiny{p}}} = -(1/2)\Delta + V$ for 
the quantum particle. $V$ is assumed to be a confining potential which grows at infinity, in particular 
$H_{\mathrm{\tiny{p}}}$ has a purely discrete spectrum. The second term is the energy of the Bose field 
with dispersion relation $\omega(k) \geq 0$. As a shorthand we set $H_{\mathrm{\tiny{f}}} = \int_{\R^d} 
\omega(k) a^*(k)a(k) dk$. $a^*(k),a(k)$ are the Bose creation and annihilation operators which satisfy 
the canonical commutation relations (CCR) $[a(k), a^*(k')] = \delta(k-k')$, $[a(k),a(k')] = 0 = [a^*(k),
a^*(k')]$. The last term in (\ref{fs}) describes a translation invariant interaction between the particle 
and the field, which is linear in the field operators. $\rh$ is the charge distribution and mollifies the 
coupling between particle and field. We assume $\rh$ to be smooth, radial and of rapid decay. $\hat\rh$ 
denotes the Fourier transform of $\rh$. We use the normalization $\int_{\R^d} \rh(x) dx = e \geq 0$; $e$ 
calibrates then the strength of the interaction. $H_{\mbox{\tiny{gN}}}$ acts on $L^2(\R^d,dq) \otimes  
{\cal F}_{\mathrm{sym}} $ with ${\cal F}_{\mathrm{sym}}$ the symmetric Fock space over $L^2(\R^d,dk)$. If 
$\int_{\R^d} |\hat\rh(k)|^2(\omega(k)^{-1} + \omega(k)^{-2}) dk < \infty$, then $H_{\mbox{\tiny{gN}}}$ is 
bounded from below and the interaction is infinitesimally $H_{\mathrm{\tiny{f}}}$-bounded. Therefore 
$H_{\mbox{\tiny{gN}}}$ is a self-adjoint operator with domain $D(H_{\mathrm{\tiny{p}}} \otimes 1) \cap 
D(1 \otimes H_{\mathrm{\tiny{f}}})$. $\rh$ introduces an ultraviolet cut-off, which may be removed by 
letting $\rh(x) \to e\delta(x)$ or, equivalently, $\hat\rh (k) \to (2\pi)^{-d/2} e$. In his famous paper 
\cite{Nel1} Nelson proved that in the case $\omega(k) = \sqrt{k^2 + m_{\rm{\tiny{b}}}^2}$, 
$m_{\rm{\tiny{b}}} > 0$, at the expense of renormalizing the energy, $H_{\mbox{\tiny{gN}}}$ stays 
well-defined in the point charge limit. Here we will keep the ultraviolet cut-off and want to study the 
infrared behaviour, which is of interest only for massless bosons, $m_{\rm{\tiny{b}}} = 0$, i.e. 
\begin{equation}
\omega(k) = |k|.
\label{ird}
\end{equation}
We call $H_{\mbox{\tiny{gN}}}$ with the dispersion relation (\ref{ird}) the massless Nelson model and 
denote the corresponding Hamiltonian by $H_{\mbox{\tiny{N}}}$. For $d = 1,2$ the Nelson Hamiltonian is 
not bounded from below, thus we require $d \geq 3$. In fact, we will mostly concentrate on dimension 
$d = 3$ for which, as proved below, the massless Nelson model is infrared divergent. 

Before stating our specific project let us explain in general terms the phenomenon of infrared divergence
in the context of Hamiltonians of the type (\ref{fs}). It is convenient to generalize somewhat and start
from
\begin{equation}
H_{\mathrm{\tiny{G}}} = K \otimes 1 + 1 \otimes H_{\mathrm{\tiny{f}}} + \int_{\R^d} \left( v(k) \otimes 
a^*(k) + v(k)^* \otimes a(k) \right) dk.
\label{gen}
\end{equation}
$K$ is a self-adjoint operator acting on the particle space $\K$ and it is assumed to have a purely 
discrete spectrum. $k \mapsto v(k)$ is a function with values in $\K$ such that $H_{\mathrm{\tiny{G}}}$ 
is bounded from below. Clearly, $H_{\mbox{\tiny{gN}}}$ is the special case where $\K = L^2(\R^d,dq)$, 
$K = H_{\mathrm{\tiny{p}}}$, $v(k) = (2\pi)^{-d/2} e^{-ik \cdot q} \hat\rh(-k)/(2\omega(k))^{1/2}$. The 
Hamiltonian $H_{\mathrm{\tiny{G}}}$ acts on $\K \otimes {\cal F}_{\mathrm{sym}}$. We say that 
$H_{\mathrm{\tiny{G}}}$ is {\it infrared divergent} if it has no ground state in $\K \otimes 
{\cal F}_{\mathrm{sym}}$. The physical ground state has then an infinite number of very low energy 
bosons, which forces a representation of the CCR different from the Fock representation. 
$H_{\mathrm{\tiny{G}}}$ should be regarded as a formal object only and one has first to construct 
the Hilbert space containing the physical ground state, and the corresponding Hamiltonian.

Ideally, one would like to decide infrared divergence directly in terms of $K$ and $v$ in (\ref{gen}). 
At present, this issue is only partially understood. G\'erard \cite{Ger} proves that if, beside some 
conditions ensuring the self-adjointness of $H$,
\begin{equation}
\int_{\R^d} \omega(k)^{-2} \|v(k) (K + 1)^{-1} \|^2 dk \; < \; \infty
\label{ger}
\end{equation}
holds, then $H_{\mathrm{\tiny{G}}}$ has a ground state in $\K \otimes {\cal F}_{\mathrm{sym}}$. In the 
particular case of (\ref{fs}), if
\begin{equation}
\int_{\R^d} |\hat\rh(k)|^2 \left(\omega (k) + \omega(k)^{-2} + \omega(k)^{-3} \right) dk \; < \; \infty,
\label{spo}
\end{equation}
then $H_{\mathrm{\tiny{gN}}}$ has a unique ground state in $L^2(\R^d,dx) \otimes {\cal F}_{\mathrm{sym}}$,  
\cite{Spo2}. Thus (\ref{spo}) has to be violated for $H_{\mathrm{\tiny{gN}}}$ to be infrared divergent. We 
note that for the massless Nelson model in three dimensions indeed
\begin{equation}
\int_{\R^3} \frac{|\hat\rh(k)|^2}{|k|^3} dk \; = \; \infty.
\end{equation}
On the other hand, by (\ref{spo}) the massless Nelson model is infrared convergent for $d\geq 4$. 

The integrals in (\ref{ger}), resp. (\ref{spo}), becoming infinite is only a necessary condition for
infrared divergence to occur. To conjecture sufficient conditions we have to rely more closely on model 
systems. The best understood model is the spin-boson Hamiltonian where $\K = \mathbb{C}^2$, $K = \sigma_x$,
$v(k) = \lambda(k)\sigma_z$, and $\sigma_x, \sigma_z$ being the Pauli matrices. Some indication comes from
the generalized spin-boson models studied in \cite{AHH}, where $v(k) = \sum_{j=1}^J B_j\lambda_j(k)$ with 
$B_j = B_j^*$ acting on $\K$. From these examples we infer that, in addition to (\ref{ger}) turning infinite, 
$v(k)$ must have non-vanishing ground state expectation for small $k$. If this happens, the model is 
divergent for {\it all} couplings $e \neq 0$. If, however, $v(0)$ has zero ground state expectation then 
there may occur cancellations making the model stay infrared convergent in spite of the integral in 
(\ref{ger}) being infinite. This is the case for the spin-boson Hamiltonian with $v(k) = \alpha \lambda(k) 
(\sigma_z + m)$ and $\int_{\R^d} |\lambda|^2/\omega^2 dk = \infty$. If $m \neq 0$, the model is infrared 
divergent for all $\alpha \neq 0$. If $m = 0$, cancellations appear and for sufficiently small $\alpha$ 
the model is infrared convergent while for large $\alpha$ it is divergent \cite{Spo2}. The same phenomenon 
occurs in the dipole approximation of the massless Nelson model in $d = 1$ where $\K = L^2(\R,dq)$, $K = 
H_{\mathrm{\tiny{p}}}$, $v(k) = q \cdot \hat\rh(k)$ \cite{OS}. If $V(q) = V(-q)$, infrared divergence 
depends on how strong the coupling is. Otherwise the model is infrared divergent for all couplings $e 
\neq 0$. In the three dimensional Nelson Hamiltonian we have
\begin{equation}
v(k) = \frac{e}{\sqrt{2|k|}} \; > \; 0
\end{equation}
for small $k$. Thus we expect, and will in fact prove, that $H_{\mbox{\tiny{N}}}$ is infrared divergent
for all couplings $e \neq 0$. 

Fr\"ohlich \cite{Fro} considers the massless Nelson model for $V = 0$ at fixed total momentum $p$ in 
which case the Hamiltonian reads
\begin{equation}
H(p) = \frac{1}{2} \left( p - \int_{\R^3} k a^*(k)a(k) dk \right)^2 + H_{\mathrm{\tiny{f}}} + 
\frac{1}{(2\pi)^{3/2}}\int_{\R^3} \frac{1}{\sqrt{2|k|}} (\hat\rh(k) a(k) + \hat\rh(-k) a^*(k)) dk 
\end{equation}
acting on ${\cal F}_{\mathrm{sym}}$. With infrared cut-off and for sufficiently small $|p|$, $H(p)$ has
a unique ground state in ${\cal F}_{\mathrm{sym}}$. Removing the infrared cut-off yields a physical 
ground state outside Fock space and a non-Fock representation of CCR. \cite{Fro} develops the 
$C^*$-algebraic approach. In contrast, in our work we follow the route of constructive quantum field 
theory by introducing a suitable probability measure on the space of classical paths. In terms of this 
path measure we can define both the physical ground state and the Hamiltonian governing the excitations
relative to the ground state. While the explicit construction will take the rest of the paper, we first
want to explain some of the particular features of the path measure for $H_{\mbox{\tiny{N}}}$. 

In the Euclidean picture we have the trajectories $t \mapsto q_t$ of the particle, resp. $t \mapsto 
\xi_t(x)$, $x \in \R^d$, of the field. To attach to them a statistical weight we define the formal 
Euclidean action by 
\begin{equation}
S_{\mathrm{p}}(\{q_t\}) = \int_\R \left(\frac{1}{2} \dot q_t^2 + V(q_t) \right) dt
\end{equation}
for the particle, and 
\begin{equation}
S_{\mathrm{f}}(\{\xi_t\}) = \int_\R\int_{\R^d} \frac{1}{2} \left((\partial_t\xi_t(x))^2 + 
(\nabla \xi_t(x))^2\right) dx dt \label{ff}
\end{equation}
for the field. From (\ref{fs}) we infer that the interaction restricted to the time interval $[-T,T]$ 
is given by
\begin{equation}
S_{\mathrm{int},T}(\{q_t,\xi_t\}) = \int_{-T}^T \int_{\R^d} \rh(x-q_t) \xi(t,x) dx dt
= \int_{-T}^T (\xi_t * \rh)(q_t) dt
\label{iiii}
\end{equation}
with $*$ denoting convolution. Thus, heuristically, the Euclidean path measure is
\begin{equation}
\frac{1}{Z} \prod_{t \in \R} dq_t e^{-S_{\mathrm{p}}(\{q_t\})} \; 
\prod_{(t,x) \in \R^{d+1}} d\xi_t(x)\; e^{-S_{\mathrm{f}}(\{\xi_t\})} \; 
e^{-S_{\mathrm{int},T}(\{q_t,\xi_t\})}.
\label{heu}
\end{equation}
The first factor above provides an a priori weight for the paths $q_t$. The corresponding path measure
is denoted by $\N^0$, compare with Section 2.1 below. The second factor provides an a priori weight for
the paths $\xi_t$. Clearly, it is a Gaussian measure which we denote by $\G$, see Section 2.2 below. In 
the spirit of the Feynman-Kac formula we define the semigroup
\begin{equation}
(F,e^{-tH}G) = \mathbb{E}_{\N^0\times\G}[F(q_0,\xi_0) G(q_t,\xi_t) e^{-\int_0^t (\xi_s*\rh)(q_s) ds}] 
\label{fk}
\end{equation}
$t \geq 0$. Then $e^{-tH}$ is a contractive semigroup on $L^2(\R^d,d\n^0) \otimes L^2({\cal S}',d\g)$, 
where $d\n^0$ and $d\g$ are the single time distributions under $\N^0$ and $\G$, respectively. $e^{-tH}$ 
is unitarily equivalent with $e^{-tH_{\mbox{\tiny{N}}}}$, $t\geq 0$, on $L^2(\R^d,dq) \otimes 
{\cal F}_{\mathrm{sym}}$.  

To construct the physical ground state and its associated Hilbert space we have to give sense to the 
path measure (\ref{heu}) in the infinite time limit $T \to \infty$. With respect to the limit measure 
(whenever it exists), the process $t \mapsto q_t, \xi_t(x)$ jointly is a time reversible stationary 
Markov process. Therefore this stochastic process has a canonically defined semigroup $T_t$, $t \geq 0$, 
self-adjoint on the Hilbert space weighted by the time $t=0$ stationary measure. Thus $T_t$ has a 
self-adjoint generator so that $T_t = e^{-tH_{\mathrm{euc}}}$, $t \geq 0$, which by definition is the 
Euclidean Hamiltonian $H_{\mathrm{euc}}$. Clearly, by construction, $H_{\mathrm{euc}} \geq 0$ and 
$H_{\mathrm{euc}} 1 = 0$, i.e. $H_{\mathrm{euc}}$ has an eigenvalue at the bottom of its spectrum. 

If $H_{\mathrm{\tiny{N}}}$ is infrared convergent, then $H_{\mathrm{euc}}$ and $H$ are unitarily 
equivalent and one may work with either of them on an equal footing. However, in the infrared divergent 
case they are not unitarily equivalent. Since $H_{\mathrm{euc}}$ accomodates the ground state, it should 
be regarded as the proper Hamiltonian of the system. 

A short outline of the paper is the following. We recall in Section 2 the basic processes leading up
to Section 3 where the definition of the joint particle-field path measure appearing in (\ref{heu}) is
given. Based on the cluster expansion in \cite{LM}, under suitable assumptions on $V$ and for sufficiently 
small $e$, we establish the existence of the infinite time measure. Thereby $H_{\mathrm{euc}}$ is well 
defined. In Section 4 we characterize infrared divergence in terms of the path measure for the interacting
system and 
prove that the massless Nelson model in $d = 3$ is infrared divergent. In Section 5 we discuss the infrared
convergent case appearing for $d \geq 4$. Finally, in Section 6 we show that $H_{\mathrm{euc}}$ has no 
spectral gap.

We remark that the quantized Maxwell field is less infrared divergent since by minimal coupling the
Bose field couples to the momentum and not to the position as in (\ref{fs}). As proved in \cite{BFS,GLL}, 
the Pauli-Fierz operator has a ground state in Fock space also in three dimensions, which suggests that 
in the case of the Maxwell field $H_{\mathrm{euc}}$ and $H$ are unitarily equivalent.

\section{Path measures for the particle and the field}

\subsection{Euclidean quantization of the particle}
On the Hilbert space $L^2(\R^d,dq)$ we define the particle Hamiltonian 
\begin{equation}
H_{\mathrm{p}} = -\frac{1}{2} \Delta + V(q).
\label{schr}
\end{equation}
The external potential $V: \R^d \to \R$ is a multiplication  operator. We consider two classes of 
potentials.
\begin{trivlist}
\item
(P1) $V$ is bounded from below, continuous, and having the asymptotics 
\begin{equation}
V(q) = C |q|^{2\alpha} + o(|q|^{2\alpha})
\label{vasy}
\end{equation}
for large $|q|$, with some constant $C > 0$ and exponent $\alpha > 1$. 

\vspace{0.3cm}
\item
(P2) $V$ is of Kato-class bounded from below as
\begin{equation}
V(q) \; \geq \; C|q|^\alpha
\end{equation}
with some constant $C > 0$ and exponent $\alpha > 0$. 
\end{trivlist}
For details on the Kato class we refer to \cite{BL}. In either case above $H_{\mathrm{p}}$ has discrete 
spectrum and a unique, strictly positive ground state $\psi_0$ lying at its bottom $E_{\mathrm{p}}$, i.e. 
$H_{\mathrm{p}}\psi_0 = E_{\mathrm{p}} \psi_0$. Associated with $H_{\mathrm{p}}$ we consider the stationary 
$P(\phi)_1$-process $q_t$ with time $t=0$ distribution (invariant measure) 
\begin{equation}
d\n^0 = \psi_0^2(q) dq, 
\end{equation}
as defined through the stochastic differential equation
\begin{equation}
dq_t = \nabla \ln \psi_0 (q_t) \; dt + dB_t
\end{equation}
where $dB_t$ denotes Brownian motion. The path measure of this process will be denoted by $\N^0$; for 
more details see \cite{BL} and references therein. (Here and henceforth we use calligraphic letters for path
measures and straight characters for time $t=0$ distributions.)
\begin{lemma}
$\N^0$-almost all paths $Q = \{q_t\}$ are continuous and for any $T > 0$ they satisfy
\begin{equation}
|q_t - q_s| \;\leq\; C |t-s|^{1/8}  \;\;\; \forall s,t \in [-T,T],
\label{pac}
\end{equation}
where $C = C(Q,T) > 0$ is independent of $s, t$. Moreover, $\N^0$-almost surely
\begin{equation}
|q_t| \;\leq\; C_1 (\ln (|t| + 1))^{1/(\alpha+1)} + C_2, \;\;\; \forall t \in \R,
\label{pas}
\end{equation}
where $C_1 > 0$ is a constant dependent only on the dimension of the space, $C_2 = C_2(Q)$, and 
$\alpha$ is the exponent appearing in (P1), resp. (P2).  
\label{lbl}
\end{lemma}
\begin{pf}
See \cite{BL}.
\end{pf}
\vspace{0.2cm}
The process $q_t$ is time-reversible. Its stochastic semigroup is generated by 
\begin{equation}
H_{\mathrm{p}}^{\mathrm{\tiny{euc}}} f = -\frac{1}{2} \Delta f - \nabla \ln \psi_0 \cdot \nabla f. 
\end{equation}
$H_{\mathrm{p}}^{\mathrm{\tiny{euc}}}$ is self-adjoint on the Hilbert space $L^2(\R^d,d\n^0)$. The 
map 
\begin{equation}
f \mapsto f/\psi_0
\end{equation}
from $L^2(\R^d,dx)$ to $L^2(\R^d,d\n^0)$ is unitary and transforms $H_{\mathrm{p}} - E_{\mathrm{p}}$ 
into $H_{\mathrm{p}}^{\mathrm{\tiny{euc}}}$.  

\subsection{Euclidean quantization of the free field}
We introduce the massless free field in $d+1$ dimensions, $d \geq 3$. For any $f \in {\cal S} 
(\R^{d+1})$, i.e. the Schwartz space over $\R^{d+1}$, we denote
\begin{equation}
\xi(f) = \int_{\R^{d+1}} f(t,x) \xi_t(x) dt dx.
\end{equation}
According to the action (\ref{ff}) the quantum field $\{ \xi (f), f \in {\cal S} (\R^{d+1})\}$ is 
described by the generalized Gaussian measure $\G$ on $\mathcal{S}'(\R^{d+1})$ with zero mean and 
covariance
\begin{eqnarray}
\cov_{\G} (\xi(f_1), \xi (f_2)) 
& = &
\mathbb{E}_{\G} [\xi(f_1) \xi(f_2)] \nonumber \\
& = & 
\frac{1}{2}\left((-\partial_ t^2 - \Delta)^{-1} f_1, f_2 \right)
\nonumber \\
& = & 
\frac{1}{2}\int_{\R} \int_{\R^d} \frac{\hat{f}_1 (k_0,k) \hat{f}_2^* (k_0,k)}{k_0^2 + k^2} dk dk_0.
\label{gc}
\end{eqnarray}
Here $\hat{f}_1, \hat{f}_2 \in \mathcal{S}(\R^{d+1})$ denote the Fourier transforms of $f_1, f_2 
\in \mathcal{S} (\R^{d+1})$. The scalar product is taken in the space $L^2(\R^{d+1}, dt dx)$. 
$\G$ is a probability measure on the space ${\cal S}'(\R^{d+1})$ taken together with its associated 
Borel $\sigma$-field \cite{DM,Min}. 

The particle trajectories are a stochastic process $q_t$ taking values in $\R^d$. Following this 
example, it will be convenient to regard the field configurations $\xi_t$ as a stochastic process 
taking values in a suitable Hilbert space to be specified below. To carry out such a construction 
first note that the covariance (\ref{gc}) can be extended to test functions of the form
\begin{equation}
f^{\vp}_t(x',t') = \vp(x') \delta (t'-t) 
\end{equation}
with $\vp \in {\cal S}(\R^d)$. Thus the random process
\begin{equation}
\xi_t (\vp) = \xi(f_t^{\vp}), \;\;\; \vp \in \mathcal{S} (\R^d), \; t \in \R, 
\label{rp}
\end{equation}
is well-defined. From (\ref{gc}) and (\ref{rp}) it follows that
\begin{equation}\label{b.n}
\mathbb{E}_{\G} [\xi_s (\vp_1) \xi_t (\vp_2)] = \frac{1}{4} \int_{\R^d} \frac{\hat{\vp}_1 (k) 
\hat{\vp}_2^* (k)}{|k|} \; e^{-|k| |s-t|}  dk.
\label{cova}
\end{equation}
Clearly, $\xi_t(\vp)$ is a stationary Gaussian process. Its invariant ($t=0$) measure $\g$ defined 
on $\mathcal{S}' (\R^d)$ is again Gaussian with mean $0$ and covariance given by (\ref{cova}) at 
$s=t$. Moreover, the process $\xi_t(\vp)$ is time-reversible \cite{GS}.
\begin{lemma}
$\{\xi_t , t \in \R\}$ is a Markov process. 
\label{l1}
\end{lemma}
\noindent
\begin{pf}
See \cite{DM,Mol}. 
\end{pf} 

\vspace{0.2cm}
Next we construct a Hilbert space $\B_D$ such that $\xi_t \in \B_D$ and $t \mapsto \xi_t$ is 
norm-continuous with probability 1. Let $D$ be the positive self-adjoint operator in the space 
$L^2 (\R^d,dx)$ given by a jointly continuous symmetric kernel 
\begin{equation}
(D \hat \xi) (k)= \int_{\R^d} D(k,k') \hat \xi(k') dk'
\label{ker}
\end{equation}
with $\ker D = \{0\}$. We introduce the real Hilbert space $\B_{D} \subset \mathcal{S}'(\R^d)$ with 
norm
\begin{equation}\label{b.p}
\|\xi\|^2_{\B_D} = \int_{\R^d \times \R^d} D (k_1,k_2) \hat{\xi} (k_1) \hat{\xi}^* (k_2) dk_1 dk_2
\end{equation}
and $\hat\xi^*(k) = \hat\xi(-k)$. Clearly, 
\begin{equation}
\mathbb{E}_{\G} \left[\|\xi_t\|^2_{\B_{D}}\right] =  \mathbb{E}_{\g} \left[\|\xi_0\|^2_{\B_{D}}\right] 
=  \frac{1}{4}\int_{\R^d} \frac{D(k,k)}{|k|} dk.
\end{equation}
Hence, if 
\begin{equation}
\int_{\R^d} \frac{D(k,k)}{|k|} dk < \infty,
\label{dk}
\end{equation}
the measure $\g$ is concentrated on the space $\B_{D}$ and the random process $\xi_t$ takes its 
values from this set; for more details see \cite{GS}. By an easy calculation we obtain that
\begin{eqnarray}
\lefteqn{
\mathbb{E}_{\G}(\|\xi_t-\xi_s\|^4_{\B_D}) = \frac{1}{4} \left(\int_{\R^d} \frac{D(k,k)}{|k|}
(1-e^{-|k| |t-s|})dk \right)^2 + } \\ &&
\hspace{4cm}
\frac{1}{2} \int_{\R^d}\int_{\R^d} \frac{D(k_1,k_2)^2}{|k_1| |k_2|} (1-e^{-|k_1| |s-t|})
(1-e^{-|k_2| |s-t|}) dk_1 dk_2.
\nonumber
\end{eqnarray}
Thus if in addition to (\ref{dk}) we assume that
\begin{equation}\label{b.s}
\int_{\R^d} D (k,k) dk < \infty \;\;\; \mbox{and} \;\;\; \int \int D(k_1,k_2)^2 dk_1 dk_2 < \infty,
\label{dkk}
\end{equation}
then, using that $|k|^{-1} (1 - e^{-|k| |s-t|}) \;\leq\; |s-t|$, we obtain
\begin{equation}
\mathbb{E}_{\G} \left[ \|\xi_t - \xi_s\|^4 \right]  \;\leq\; C \; |t-s|^2.
\label{cont}
\end{equation}
From (\ref{cont}) and Kolmogorov's criterion of continuity of random processes 
(see \cite{GS,Shi}) we conclude
\begin{lemma}
Under conditions (\ref{dk}) and (\ref{dkk}) almost all paths of the process $\xi_t$ are continuous 
in the metric of $\B_{D}$. Moreover, for any finite interval $T > 0$ these paths have the property 
that 
\begin{equation}
\|\xi_s - \xi_t\|_{\B_{D}} \;\leq\; C \; |t-s|^{1/8}, \;\;\; \forall s,t \in [-T,T]
\label{pc}
\end{equation}
where $C = C(\{\xi_t\},T)$ is independent of $s,t$.
\label{cl}
\end{lemma}
Lemma \ref{cl} states that with probability 1 the process $\xi_t$ is realized on $C(\R,\B_D)$. We 
denote the corresponding Gaussian path measure again by $\G$.  Since the stationary Markov process 
$\xi_t$ is reversible, its stochastic semigroup $e^{-t H_{\mathrm{f}}^{\mathrm{\tiny{euc}}}}$ acting 
on $L^2(\B_D,d\g)$ is self-adjoint and $H_{\mathrm{f}}^{\mathrm{\tiny{euc}}}$ is the generator of 
this semigroup. 
\begin{lemma}
There exists a unitary map ${\cal F}_{\mathrm{sym}}\to L^2(\B_D,d\g)$ transforming the free
field Fock space Hamiltonian $H_{\mathrm{f}}$ into $H_{\mathrm{f}}^{\mathrm{\tiny{euc}}}$. 
\end{lemma}
This map can be constructed by using the It\^o-Wick transformation, see \cite{Sim}.

For concreteness we give an explicit form of the kernel $D (k_1, k_2)$ which satisfies (\ref{dk}) 
and (\ref{dkk}) above. Consider the operator
\begin{equation}
\bar D = (- \Delta_k + |k|^2)^{-(d+1)}. 
\end{equation}
The eigenvalues of this operator are 
\begin{equation}
\lambda_n = \frac{1}{(n_1 + ... + n_d + d/2)^{d+1}}, 
\end{equation}
with multi-index $n = (n_1,...,n_d)$ of integer entries $n_i \geq 0$, $i=1,...,d$. Hence $\bar D$ 
is a nuclear and positive operator, and its kernel $\bar D(k_1, k_2)$ is jointly continuous 
\cite{Ber}. Let $|k|_1$ denote the function
\begin{equation}
|k|_1 =
\left\{
\begin{array}{ll}
\vspace{0.2cm}
|k|, & \mbox{if $|k| < 1$,} 
\nonumber \\
1, & \mbox{otherwise.}
\end{array} \right.
\end{equation}
We set
\begin{equation}
D(k_1,k_2) = |k_1|_1 \bar{D} (k_1, k_2) |k_2|_1.
\end{equation}
It is then easily seen that this kernel satisfies conditions (\ref{dk}) and (\ref{dkk}).

For later use we also note that the level sets 
\begin{equation}
K_C = \{\xi \in \B_D: (D^{-1/2} \xi,\xi)_{\B_D} \; \leq \; C \}
\label{kac}
\end{equation}
are compact in $\B_D$. 

\section{Path measure for the interacting system}
In the previous section we defined the stochastic processes $q_t$ and $\xi_t$ governing the 
particle, resp. the free field trajectories. The joint field-particle process on the space 
$C(\R,\R^d \times \B_D)$ is then given by the path measure $\p^0 = \N^0 \times \G$. In order 
to implement the interaction as in (\ref{heu}) we take $\p^0$ as reference process and modify 
it with a density given by the exponential of $S_{\mathrm{int}}$, compare with (\ref{iiii}),
\begin{equation}
d\p_T  = \frac{1}{Z_T} \exp\left(-\int_{-T}^T (\xi_t * \rh)(q_t) dt\right) d\p^0,
\label{fpp}
\end{equation}
where 
\begin{equation}
Z_T = \int \exp\left(-\int_{-T}^T  (\xi_t * \rh)(q_t) dt \right) d\p^0 
\end{equation}
is the normalizing partition function. Since $\G$ is a Gaussian measure, by an easy computation 
we obtain
\begin{equation}
Z_T = \int \exp\left(\frac{1}{8}\int_{-T}^T \int_{-T}^T dt ds \int_{\R^d} \frac{|\hat\rh(k)|^2}
{|k|} \cos (k \cdot (q_t-q_s)) \; e^{-|k| |t-s|} dk \right) d\N^0.
\end{equation}
This integral is well-defined for every $T > 0$. 

We will be interested whether the sequence of measures $\p_T$ has a limit as $T \to \infty$. We 
use the following notion of convergence. Let $M$ be a metric space, and $C(\R,M)$ the space of 
continuous paths $\{X_t\}$ with values in $M$. For any interval $I_T = [-T,T] \subset \R$ let 
$\M_T \subset \C$ be a sub-$\sigma$ field of the Borel $\sigma$-field $\C$ generated by the 
evaluations $\{X_t: t \in I_T\}$, i.e. $\M_T = \sigma(X_t, t \in I_T)$. We say that a sequence 
of probability measures $\{\mu_n\}$ on $C(\R,M)$ converges locally weakly to the probability 
measure $\mu$ if for any $0 \leq T < \infty$ the restrictions $\mu_n|_{\M_T}$ converge weakly to 
the measure $\mu_{\M_T}$, see \cite{GS}. We have then the following result.
\begin{theorem}
Let $d \geq 3$, $0 < e \leq e^*$, with sufficiently small $e^* > 0$, and $V$ satisfy (P1). Then
there exists the local weak limit $\p_T \to \p$ as $T \to \infty$. $\p$ is a probability measure 
of a stationary reversible Markov process on the path space $C(\R,\R^d \times \B_{D})$. Moreover, 
the stationary distribution $\pp$ of $\p$ can be obtained as the weak limit $\pp = \lim_{T \to 
\infty} \pp_T$, where $\pp_T$ is the distribution of $\{q_0,\xi_0\}$ under $\p_T$. 
\label{t1}
\end{theorem}
\noindent
\begin{pf}
For any fixed path $ Q = \{q_t: t \in \R\} \in C(\R,\R^d)$ consider the conditional 
distribution 
\begin{equation}
\p^{Q}_T = \p_T(\;\cdot\;| \{q_t\} = Q)
\end{equation}
viewed as a probability measure on $C(\R,\B_{D})$. Then the distribution $\p_T$ on $C(\R,\R^d 
\times \B_{D})$ can be written as
\begin{equation}
\p_T(S \times A) = \int_S \p_T^{ Q} (A) d\N_T(Q), \;\;\; \forall T > 0
\label{proj}
\end{equation}
where $A \subset C(\R,\B_{D})$ and $S \subset C(\R,\R^d)$, and $\N_T$ is the $Q$-marginal of 
$\p_T$. Since the $\xi_t$ process is Gaussian and the coupling is linear in $\xi_t$, this 
marginal can be computed explicitly and is given by
\begin{equation}
d\N_T = \frac{1}{Z_T} \exp\left(-\int_{-T}^T  \int_{-T}^T W(q_s-q_t,s-t) ds dt\right) d \N^0, 
\end{equation}
where
\begin{equation}
W(q,t) = -\frac{1}{8} \int_{\R^d} \frac{|\hat\rh(k)|^2}{|k|} \cos (k\cdot q) \;  e^{-|k| |t|} dk.
\label{w}
\end{equation}
${\cal N}_T$ is a Gibbs measure for the finite interval $[-T,T]$. The reversible diffusion process
${\cal N}^0$ is the reference measure, and $W$ is the (effective) pair interaction potential. We 
denote the distribution of $q_0$ under $\N_T$ by $\n_T$. 

We use identity (\ref{proj}) to show the existence of the $T \to \infty$ limit of $\p_T$. We 
will do this by examining separately the limits of $\p_T^{Q}$ and $\N_T$, starting with the 
former measure. Since the action $\int_{-T}^T (\xi_t * \rh)(q_t) dt$ is linear in $\xi_t$, 
$\p^{Q}_T$ is a Gaussian measure with covariance given by (\ref{gc}) and mean 
\begin{equation}
\mathbb{E}_{\p^{Q}_T} [\xi_t(\vp)] = \int_{\R^d} \vp(x) g^t_T(x;Q) dx, \;\;\; \vp \in 
\s(\R^d),
\end{equation}
where
\begin{eqnarray}
g^t_T(x;Q) 
& = &
- \frac{1}{(2\pi)^{d/2}} \int_{-T}^T \int_{\R^d} \frac{\hat\rh(k)}{4|k|} 
\cos (k \cdot(x-q_\tau)) \; 
e^{-|k| |t-\tau|} dk d\tau \label{g} \\
& = & \frac{1}{(2\pi)^{d/2}} \int_{\R^d} e^{ik\cdot x} \; \hat g^t_T(k;Q) dk \nonumber 
\end{eqnarray}
and
\begin{equation}
\hat g^t_T(k;Q) = - \frac{\hat \rh(k)}{4|k|} \int_{-T}^T e^{-i k\cdot q_\tau} 
e^{-|k| |t-\tau|} d\tau.
\label{gee}
\end{equation}
This function has the properties
\begin{eqnarray}
&& |\hat g^t_T(k;Q)| \;\leq\; \frac{|\hat\rh(k)|}{2|k|^2} \label{not-s}\\
&& |\hat g^t_T(k;Q) - \hat g^s_T(k;Q)| \;\leq\; \frac{C |\hat\rh(k)|}{2|k|} \; |t-s| 
\label{t-s} 
\end{eqnarray}
for any $s,t \in \R$, $|s-t| < 1$, and with some constant $C > 0$. Furthermore, we have that 
for any $t$ and $Q$
\begin{equation}
\lim_{T\to\infty} \hat g^t_T(k;Q) = \hat g^t(k;Q) 
\label{limg}
\end{equation}
exists in the norm topology of $\B_{D}$, uniformly with respect to $Q$ and $t$ in every 
bounded interval. The limit of (\ref{g}) is given explicitly by
\begin{equation}
g^t(x;Q) = - \frac{1}{(2\pi)^{d/2}} \int _{-\infty}^\infty \int_{\R^d} 
\frac{\hat\rh(k)}{4|k|} \cos (k\cdot (x-q_\tau)) \; e^{-|k| |t-\tau|} dk d\tau.
\end{equation}
From (\ref{t-s}) it follows that $\hat g^t_T (k;Q)$ as a function of $t$ belongs to 
$C(\R,\B_{D})$ and 
\begin{equation}
\sup_t\|\hat g^t_T (\;\cdot\;;Q)\|_{\B_{D}} < \infty
\label{ub}
\end{equation}
uniformly in $Q$ and $T \leq \infty$. Moreover, as follows from (\ref{not-s}), for arbitrary 
$Q$, $T$, and $t$, $\hat g^t_T(\cdot, Q) \in K_C$ for some suitable $C > 0$, with $K_C$ defined 
by (\ref{kac}). Thus by (\ref{t-s}) and (\ref{ub}) the set of restrictions $g^t_T(\cdot,Q)$, 
$|t| \leq R$ to any interval $[-R,R]$, $\forall Q \in C(\R,\R^d)$, $\forall T \leq \infty$, is 
a compact set in $C([-R,R],\B_D)$. Hence it follows that the restrictions $\p^Q_T|_{\M_R}$ form 
a weakly compact set of measures. 

The characteristic functional of the Gaussian measure $\p_T^{Q}$ is
\begin{eqnarray*}
\chi_T^{Q} (f) 
& = &
\mathbb{E}_{\p^Q_T} [e^{i\xi(f)}] \\
& = & 
\exp \left (-1/8 \int_{-\infty}^\infty dt \int_{-\infty}^\infty ds \int_{\R^d} 
\frac{\hat f(s,k) {\hat f}^*(t,k)}{|k|} e^{-|k| |t-s|} dk \right) \times \\
&\times&
\exp \left (i \int_{-\infty}^\infty dt \int_{\R^d} \hat g^t_T(k;Q) {\hat f}(t,k) dk \right).
\end{eqnarray*}
From (\ref{limg}) it follows that $\chi_T^{Q} (f) \to \chi^{Q} (f)$ as $T \to \infty$,
where $\chi^{Q} (f)$ is the characteristic functional of the Gaussian measure $\p^{Q}$ 
with covariance (\ref{cova}) and mean 
\begin{equation}
\mathbb{E}_{\p^{Q}} [f^{\vp}_t] = \int_{\R^d} \vp(x) g^t(x;Q) dx.
\end{equation}
Thus from the convergence of characteristic functionals $\chi^Q_T$ and the weak compactness 
of the restrictions $\p^Q_T|_{\M_R}$ the weak local convergence $\p^Q_T \to \p^Q$ as $T \to 
\infty$ follows for any $Q \in C(\R,\R^d)$, see \cite{GS}.

In order to complete our argument we need to control the $\N_T \to \N$ limit. This problem 
has been investigated in \cite{LM} by a cluster expansion technique and some of the results 
relevant in our context are quoted in Theorem \ref{t2} below. In particular, the local weak 
limit $\N_T \to \N$ exists, as stated under (1) in Theorem \ref{t2}. This then completes the 
proof of the first claim. 

Since $\p_T$ has the Markov property and is reversible, the limit distribution $\p$ has 
the same properties. Stationarity of $\p$ follows by point (2) of Theorem \ref{t2} and the 
obvious equality
\begin{equation}
d\p^{\theta_s Q}(\xi_{t-s}) = d\p^{Q} (\xi_t)
\end{equation}
$\theta_s$ being the time-shift acting on paths. 

The existence of the weak limit $\pp = \lim_{T \to \infty} \pp_T$ follows by the argument  
above by borrowing statement (3) of Theorem \ref{t2} below. This completes the proof of the 
theorem. 
\end{pf}
 
We conclude this section by stating some facts on $\N$ and related measures. Results on the 
existence and uniqueness of Gibbs measures relative to Brownian motion are available for some 
time \cite{OS}, and a more general framework has been considered more recently in \cite{LM}. In 
the following theorem we list some results taken from \cite{LM} to be used below adapted for 
the case of Nelson's model. For details we refer to this paper, in particular for results on 
the uniqueness of $\N$ which we do not address here.      
\begin{theorem}
For $d \geq 3$, sufficiently small $e > 0$, and potentials satisfying (P1) the following 
properties hold:
\begin{enumerate}
\item
There exists the weak local limit $\N_T \to \N$ as $T \to \infty$. 
\item
The measure $\N$ is invariant with respect to time shifts and time reflections.
\item
The $t = 0$ distributions $\n_T$, resp. $\n$, of the measures $\N_T$, resp. of $\N$, are 
absolutely continuous with respect to $\n^0$ 
and
\begin{equation}
\frac{d\n_T}{d\n^0} (q_0) \to \frac{d\n}{d\n^0} (q_0) \;\;\; \mbox{as} \;\;\; T \to \infty
\end{equation}
pointwise. Moreover, there exists a constant $c > 0$ such that 
\begin{equation}
\frac{1}{c} \;\leq\; \frac{d\n_T}{d\n^0}(q_0) \;\leq\; c,
\label{abco}
\end{equation}
uniformly in $T$ and $q_0$.
\item
For $\n^0$-almost all $q$ the conditional distribution $\N_T(\;\cdot\;| q_0 = q)$ converges 
weakly locally to $\N(\;\cdot\;|q)$.
\item
$\N$-almost all paths $Q = \{q_t\} \in C(\R,\R^d)$ have the property that for all $t \in \R$
\begin{equation}
|q_t| \;\leq\; C_1 (\ln (|t| + 1))^{1/(\alpha+1)} + C_2(Q)
\label{typp}
\end{equation}
where $C_1 > 0$ is a constant, and $C_2$ is a function of the path.
\item
For any bounded functions $F_1,F_2$ on $\R^d$ the following estimate holds on their covariance:
\begin{equation}
\mathbb{E}_{\N}[F_1(q_s)F_2(q_t)] - \mathbb{E}_{\N}[F_1(q_s)] \mathbb{E}_{\N} [F_2(q_t)]  
\;\leq\; C \; \frac{\sup|F_1| \sup|F_2|}{|s-t|^\gamma+1}
\label{deca}
\end{equation}
with some $\gamma > 0$ and a suitable prefactor $C > 0$.
\end{enumerate}
\label{t2}
\end{theorem}

\section{The infrared divergent case: $d=3$}
After having introduced the path measures for the interacting system, we turn now to compare 
the Fock quantization and the Euclidean quantization of the interacting system. The first one 
is constructed through the Feynman-Kac formula,
\begin{equation}
\mathbb{E}_{\p^0}[F(q_0,\xi_0) G(q_t,\xi_t) e^{-\int_0^t (\xi_s*\rh)(q_s) ds}] = 
(F,e^{-tH}G)_{\h^0}, 
\;\;\; t \geq 0,
\label{fkh}
\end{equation}
which defines the semigroup $\exp(-tH)$ on $\h^0 = L^2(\R^d \times \B_{D}, d\pp^0)$. Its 
generator, the Hamiltonian $H$, is unitarily equivalent with $H_{\mathrm{\tiny{N}}}$. For the 
Euclidean quantization we start with the semigroup $T_t$ associated with the time reversible 
Markov stochastic process $\{q_t,\xi_t\}$ distributed according to $\p$ as defined through
\begin{equation}
\mathbb{E}_{\p}[F(q_0,\xi_0) G(q_{t},\xi_{t})] = (F,T_{t} G)_{\h}, \;\;\; t \geq 0, 
\label{euc}
\end{equation}
on the Hilbert space $\h = L^2(\R^d \times \B_{D}, d\pp)$. $T_t$ is a symmetric contractive 
semigroup. Hence there exists a self-adjoint semibounded operator $H_{\mbox{\tiny{euc}}}$ 
generating it, i.e. $T_t = \exp(-tH_{\mbox{\tiny{euc}}})$, which by definition will be viewed 
as the Hamiltonian of the system obtained in Euclidean quantization. Note that the constant 
function 1 is the ground state of $H_{\mbox{\tiny{euc}}}$. Since $T_t$ is positivity improving, 
this ground state is unique in $\h$. 

In this and the following section we investigate whether the Hamiltonians $H$ and 
$H_{\mbox{\tiny{euc}}}$ are unitarily equivalent. We start by first characterizing infrared 
divergence in terms of both the approximate ground state and the density of the path measure 
with respect to the reference measure. We use throughout this paper the following 
\begin{definition}
The Hamiltonian $H$ is called {\rm infrared divergent} if it has no ground state in $\h^0$. 
\end{definition}

Let $\{ P_\lambda \}$ be the family of spectral projections for the self-adjoint operator $H$
bounded from below, and denote $d\sigma_1(\lambda) = d(1,P_\lambda 1)_{\h^0}$, the spectral 
measure for $1 \in \h^0$. Then 
\begin{equation}
E_0 \;\leq\; E_0' := \inf \supp d\sigma_1(\lambda).
\label{e0'}
\end{equation}
$E_0$ is the bottom of the spectrum of $H$. We also define the approximate ground state 
\begin{equation}
\Psi_T = \frac{e^{-T(H-E_0')}{\mathrm 1}}{\|e^{-T(H-E_0')} {\mathrm 1}\|}.
\end{equation}

\begin{theorem}
Suppose $V$ satisfies condition (P1) and the Gibbs measure $\N$ exists. Then
the following statements are equivalent:
\begin{trivlist}
\item
(1) $H$ is infrared divergent.
\item
(2) $\lim_{T\to\infty} (1,\Psi_T) = 0$.
\item
(3) $\pp$ is singular with respect to $\pp^0$.
\end{trivlist}
\label{tir}
\end{theorem}
\begin{lemma}
We have the following two cases:
\begin{trivlist}
\item
(i) If $\limsup_{T\to\infty} (1,\Psi_T) > 0$, then $\sigma_1(\{E_0'\}) > 0$ and the limit
\begin{equation}
\lim_{T \to \infty} \Psi_T := \Psi.
\label{s1}
\end{equation}
exists. Moreover, $\Psi > 0$, $E_0' = E_0$, and $\Psi$ is the unique ground state of 
$H$ at eigenvalue $E_0$. 
\item
(ii) If $\limsup_{T\to\infty} (1,\Psi_T) = 0$, then $\sigma_1(\{E_0'\}) = 0$ and
\begin{equation}
\lim_{T\to \infty} \Psi_T = 0.
\label{s3}
\end{equation}
Moreover, $E_0' = E_0$ and $H$ has no ground state in $\h^0$. 
\end{trivlist}
\label{lem}
\end{lemma}

\vspace{0.2cm}
\begin{pf}
(i) By the spectral theorem we have 
\begin{equation}
(1,\Psi_T) = \frac{\int_{E_0'}^\infty e^{-(\lambda - E_0')T} d\sigma_1(\lambda)}
{\left(\int_{E_0'}^\infty e^{-2(\lambda - E_0')T} d\sigma_1(\lambda)\right)^{1/2}}.
\label{spec}
\end{equation}
Take some $a > E_0'$. Then 
\begin{equation}
\frac{ \int_{E_0'}^\infty e^{-(\lambda - E_0')T} d\sigma_1(\lambda)}
{\left(\int_{E_0'}^\infty e^{-2(\lambda - E_0')T} d\sigma_1(\lambda)\right)^{1/2}} \;\leq \;
\frac{\int_{E_0'}^a e^{-(\lambda - E_0')T} d\sigma_1(\lambda) + 
      \int_a^\infty e^{-(\lambda - E_0')T} d\sigma_1(\lambda)}
     {\left(\int_{E_0'}^a e^{-2(\lambda - E_0')T} d\sigma_1(\lambda)\right)^{1/2}}. 
\label{l2l}
\end{equation}
Schwartz's inequality gives
\begin{equation}
\int_{E_0'}^a e^{-(\lambda - E_0')T} d\sigma_1(\lambda) \leq 
\left(\int_{E_0'}^a e^{-2(\lambda - E_0')T} d\sigma_1(\lambda)\right)^{1/2} 
\left(\sigma_1([E_0',a])\right)^{1/2}.
\label{scw}
\end{equation}
Hence (\ref{l2l}) further becomes   
\begin{eqnarray}
\frac{\int_{E_0'}^\infty e^{-(\lambda - E_0')T} d\sigma_1(\lambda)}
{\left(\int_{E_0'}^\infty e^{-2(\lambda - E_0')T} d\sigma_1(\lambda)\right)^{1/2}} 
& \leq &
\left(\sigma_1([E_0',a])\right)^{1/2} + \frac{e^{-(a-E_0')t}}{\left(\int_{E_0'}^a 
e^{-2(\lambda - E_0')T} d\sigma_1(\lambda)\right)^{1/2}} \nonumber \\
& = &
\left(\sigma_1([E_0',a])\right)^{1/2} + 
\frac{1}{\left(\int_{E_0'}^a e^{-2(\lambda - a)T} d\sigma_1(\lambda)\right)^{1/2}}. 
\nonumber
\end{eqnarray}
Since $E_0' = \inf \supp d\sigma_1$, the denominator diverges for $T\to\infty$ and 
\begin{equation}
0 < \limsup_{T\to\infty} (1,\Psi_T) \leq \left(\sigma_1([E_0',a])\right)^{1/2}.
\end{equation}
Taking $a \to E_0'$, this implies $\sigma_1(\{E_0'\}) > 0$. Again, by the spectral theorem 
the limit (\ref{s1}) exists and $\Psi \geq 0$, since $e^{-tH}$ is positivity preserving
\cite{BFS}. $\Psi$ is an eigenfunction of $H$ at eigenvalue $E_0'$. Since $e^{-tH}$ is 
positivity improving \cite{BFS}, $\Psi > 0$ and therefore $E_0 = E_0'$ by the uniqueness 
part of the Perron-Frobenius theorem \cite{RS}. 

(ii) By contraposition take $\sigma_1(\{E_0'\}) > 0$. Then for $T\to\infty$ the numerator 
of (\ref{spec}) converges to $\sigma_1(\{E_0'\})$ and the denominator converges to
$\sqrt{\sigma_1(\{E_0'\})}$, which contradicts $\lim_{T\to\infty} (1,\Psi_T) = 0$. 

We finally prove that $E_0' = E_0$. We have that
\begin{equation}
\inf \supp d\sigma_1 (\lambda) = - \lim_{T\to\infty} \frac{1}{T} \log (1,e^{-TH}1) = E_0'.
\end{equation}
By an easy computation, for the set of functions 
\begin{equation}
0 < c^- \leq f \leq c^+ < \infty,
\label{cfc}
\end{equation}
with some constants $c^-,c^+$, we also have
\begin{equation}
\inf \supp d\sigma_f (\lambda) = - \lim_{T\to\infty} \frac{1}{T} \log (f,e^{-TH}f) = E_0'.
\end{equation}
The linear span of the set of functions defined by (\ref{cfc}) is dense in $\h$. Indeed, take 
$f$ as specific combinations of characteristic functions, $f = 1_A + \epsilon 1_{A^c}$ with 
arbitrary sets $A \subset \h$ and some $\epsilon > 0$; this set is dense in $\h$. If $E_0' > 
E_0$ then $P_{\{\lambda < E_0'\}}\h \neq \emptyset$, and the linear span of the set (\ref{cfc}) 
would not be dense in $\h$. Thus $E_0' = E_0$. 

Note that as our proof shows, actually also the limit $\lim_{T\to\infty}(1,\Psi_T)$ exists.
\end{pf}

\begin{lemma}
We have the following two cases:
\begin{trivlist}
\item
(i) If $\pp$ is absolutely continuous with respect to $\pp^0$ then $\lim_{T\to\infty}
(1,\Psi_T) > 0$ and
\begin{equation}
\frac{d\pp}{d\pp^0} = \lim_{T\to\infty} \frac{d\pp_T}{d\pp^0} = \lim_{T\to\infty} \Psi^2_T =
\Psi^2.
\label{psip}
\end{equation}
\item
(ii) If $\pp$ is singular with respect to $\pp^0$, then $\lim_{T\to\infty} (1,\Psi_T) = 0$
and
\begin{equation}
\lim_{T\to\infty} (1,\Psi_T) = \lim_{T\to\infty} \mathbb{E}_{\pp^0} 
\left[ \left( \frac{d\pp_T}{d\pp^0} \right)^{1/2} \right].
\end{equation}
\end{trivlist}
\label{lemm}
\end{lemma}

\vspace{0.2cm}
\noindent
{\it Remark:} As follows from Lemma \ref{lem} only two cases can occur: either (i) $\pp$ is
absolutely continuous with respect to $\pp^0$, or (ii) $\pp$ is singular with respect to 
$\pp^0$.

\vspace{0.2cm}
\noindent
\begin{pf}
(i) 
Note that for any finite $T>0$ the measure $\pp_T$ is absolutely continuous with respect to 
$\pp^0$, and
\begin{equation}
\frac{d\pp_T}{d\pp^0} (q,\xi) = \frac{Z^-_T(q,\xi) Z^+_T(q,\xi)}{Z_T} 
\label{kkee}
\end{equation}
where
\begin{equation}
Z^+_T(q,\xi) = \mathbb{E}_{\p^0} [e^{-\int_0^T  (\xi_t * \rh)  (q_t) dt} |q_0=q,\xi_0=\xi]. 
\label{kke}
\end{equation}
$Z^-_T$ is defined by changing $\int_0^T$ for $\int_{-T}^0$ in the exponent, and obviously 
$Z^-_T = Z^+_T$. By applying the Feynman-Kac formula we obtain
\begin{equation}
Z^+_T(q,\xi) = (e^{-TH} 1)(q,\xi)
\label{ekke}
\end{equation}
and 
\begin{equation}
Z_T = \|e^{-TH} 1\|^2
\end{equation}
in the norm of $L^2(\R^d \times \B_{D}, d\pp^0)$. Thus
\begin{equation}
\frac{d\pp_T}{d\pp^0} (q,\xi) = \frac{\left((e^{-TH} 1)(q,\xi)\right)^2}{\|e^{-TH} 1\|^2}.
\label{sqs}
\end{equation}
Since we have $\pp_T \to \pp$ as $T \to \infty$, (\ref{sqs}) implies (\ref{psip}).

(ii) By using Schwartz's inequality and $F$ given by (\ref{F}) below, we get
\begin{equation}
(\Psi_T,1) = \int \Psi_T d\pp^0 = \int \Psi_T F^{1/2} F^{-1/2} d\pp^0 \leq
\left( \int \Psi_T^2 F d\pp^0\right)^{1/2} \left(\int F^{-1} d\pp^0\right)^{1/2}. 
\end{equation}
By Lemma \ref{ldiv} below the first integral at the right hand side above converges to 
zero as $T\to\infty$ and the second integral is bounded, hence the claim follows. 
\end{pf}

\vspace{0.3cm}
\noindent
{\it Proof of Theorem \ref{tir}:} We proceed along the scheme (3) $\Rightarrow$ (2) 
$\Rightarrow$ (1) $\Rightarrow$ (3). (3) $\Rightarrow$ (2) is given by Lemma \ref{lem} (ii), 
(2) $\Rightarrow$ (1) comes about by Lemma \ref{lemm} (ii). 

To show (1) $\Rightarrow$ (3) suppose that $\pp$ is not singular with respect to $\pp^0$. Then, 
by the above Remark $\pp$ must be absolutely continuous with respect to $\pp^0$. If, however, 
$\pp \ll \pp^0$, then by Lemma \ref{lemm} (ii) we have that there is $l > 0$ such that 
$\lim_{T\to\infty} (1,\Psi_T) = l$ and $d\pp/d\pp^0 = \Psi^2$. Moreover, by Lemma \ref{lem} (i) 
we have that $\Psi = \lim_{T\to \infty} \Psi_T > 0$ and $\Psi$ is the ground state of $H$. This 
is a contradiction proving (1) $\Rightarrow$ (3). $\Box$

\begin{theorem}
For $d = 3$, (P1)-potentials and $0 < e \leq e^*$ $H$ has no ground state in $\h^0$ (and hence 
$H_{\mbox{\tiny{N}}}$ has no ground state in $L^2(\R^d,dq) \otimes {\cal F}_{\mathrm{sym}}$). In 
particular, $H$ and $H_{\mathrm{\tiny{euc}}}$ are not unitarily equivalent. 
\label{tdiv}
\end{theorem}
We expect that the restriction to weak coupling is of technical nature only. 

\begin{lemma}
For $d=3$ and $0 < e \leq e^*$ $\pp$ is singular with respect to $\pp^0$. 
\label{ldiv}
\end{lemma}
\begin{pf}
Clearly, it suffices to find a function $F$ such that $F$ is almost surely strictly positive 
with respect to the free measure $\pp^0$ and $\mathbb{E}_{\pp}[F] = 0$. We choose
\begin{equation}
F(q,\xi) = \exp \left( \int_{\R^3} s(x) \xi(x) dx \right) 
\label{F}
\end{equation}
and $s$ satisfying
\begin{equation}
\|s\|^2 := \int_{\R^3} \frac{|\hat s(k)|^2}{|k|} dk \; < \; \infty.
\label{nor}
\end{equation}
The function $F$ is defined and positive for $\g$-almost all $\xi \in \B_{D}$ (see 
\cite{Lif}). We have the identity
\begin{equation}
\mathbb{E}_{\pp_T}[F] = \mathbb{E}_{\N_T} \left[\mathbb{E}_{\pp_T^{ Q}}[F] \right] 
\end{equation}
According to our standard notation $\pp_T^{Q}$ is the time $t=0$ distribution of $\p_T^Q$. 
$\pp^Q_T$  is absolutely continuous with respect to $\g$ for any $Q$ and its density is given by
\begin{equation}
\frac{d\pp_T^{Q}}{d\g} = \exp\left(\int_{\R^3} \xi(x) m_T(x;Q) dx - 2 \int_{\R^3} 
|\hat g^0_T (k;Q)|^2 |k| dk \right)
\label{den}
\end{equation}
with
\begin{equation}
m_T(x;Q) = \frac{1}{(2\pi)^{d/2}} \int_{\R^3} \hat m_T(k;Q) e^{ik \cdot x} dk 
\label{mT}
\end{equation}
and
\begin{equation}
\hat m_T(k;Q) = 4|k| \hat g_T^0(k;Q) = - \hat \rh(k) \int_{-T}^T e^{-ik \cdot q_\tau} 
e^{-|k| |\tau|} d\tau.
\end{equation}
Thus we obtain 
\begin{equation}
\mathbb{E}_{\pp_T^Q} [F] = \exp\left(\int_{\R^3} \hat s(k) \hat g^0_T(k;Q) dk + 
\frac{1}{8} \|s\|^2\right).
\label{co1}
\end{equation}
Since $\hat\rh(0) = e >0$, there exists some $k^* > 0$ such that $\hat \rh(k) > 0$ 
whenever $0 < |k| < k^*$. We choose then the Fourier transform of $s$ as
\begin{equation}
\hat s(k) =
\left\{
\begin{array}{ll}
\vspace{0.2cm}
\frac{1}{\hat \rh(k)} \int_{T^*}^\infty \frac{e^{-|k| |t|}}{\ln t (\ln\ln t)^\zeta} dt, 
& \mbox{if $|k| < k^*$} 
\nonumber \\
0, & \mbox{otherwise}
\end{array} \right.
\end{equation}
with a suitable constant $T^*$ such that $\ln T^* > 1$, and exponent $0<\zeta<1$. First 
we have to show that (\ref{nor}) is satisfied,
\begin{eqnarray}
\int_{\R^3} \frac{|\hat s(k)|^2}{|k|} dk \nonumber
&\leq& 
C \int_0^{k^*} k dk \int_{T^*}^\infty \int_{T^*}^\infty \frac{e^{-|k|(s+t)} ds dt}
{\ln t \ln s (\ln\ln t \ln \ln s)^\zeta} \nonumber \\
&=&
C \int_{T^*}^\infty \int_{T^*}^\infty \frac{1}{(t+s)^2} \frac{dtds}{\ln t \ln s 
(\ln\ln t)^\zeta (\ln\ln s)^\zeta} \nonumber \\
&=&
2C \int\int_{T^* \leq s \leq t < \infty} \frac{t/s}{(1+t/s)^2} \frac{dt}{t\ln t 
(\ln\ln t)^\zeta} \frac{ds}{s\ln s (\ln\ln s)^\zeta} \nonumber \\
&=&
2C \int_0^\infty \frac{e^w dw}{(1+e^w)^2} \int_{u^*}^\infty \frac{dv}{v (\ln v)^\zeta 
(v+w)(\ln (v+w))^\zeta}
\nonumber \\
&\leq&
2C \int_0^\infty e^{-w}dw \int_{u^*}^\infty \frac{dv}{v^2(\ln v)^{2\zeta}} < \infty
\label{aaa}
\end{eqnarray}
with $u = \ln t$, $v = \ln s$, $w = u-v$, $u^* = \ln T^*$ above. Now we write
\begin{eqnarray}
\int_{\R^3} \hat s(k) \hat g^0_T (k; Q) d k 
& = &
-\frac{1}{2} \int_{|k| < k^*} \frac{dk}{|k|} \int_{-T}^T e^{-i(k, q_s)} e^{-|k| |s|} ds 
\int_{T^*}^\infty \frac {e^{-|k| |t|}}{\ln t (\ln \ln t)^\zeta} dt  \nonumber \\
& = &
-2\pi \int_{-T}^T ds \int_{T^*}^\infty dt \int_0^{k^*} \frac{\sin k|q_s| \; 
e^{-k(|s| + |t|)}}{|q_s| \ln t (\ln \ln t)^\zeta} dk \nonumber \\
& = &
-4\pi  \int_{-T}^T ds \int_{T^*}^\infty dt \frac{1}{[|q_s|^2 + (|s| + |t|)^2] \ln t 
(\ln\ln t)^\zeta} \nonumber\\
& - &
4\pi \int_{-T}^T ds \int_{T^*}^\infty dt \frac{e^{-|k^*|(|s|+|t|)}\cos(k^*|q_s|)}
{[|q_s|^2 + (|s|+|t|)^2]\ln t (\ln\ln t)^\zeta}
\nonumber \\
& + &
4\pi \int_{-T}^T ds \int_{T^*}^\infty dt \frac{e^{-|k^*|(|s|+|t|)} 
\sin (k^*|q_s|/|q_s|)(|t|+|s|)}{[|q_s|^2 + (|s|+|t|)^2]\ln t (\ln\ln t)^\zeta}.  
\label{bbb} 
\end{eqnarray}
The second and third integrals in the latter expression converge for all $ Q$ in the $T 
\to \infty$ limit. Take now the first integral. By (\ref{typp}) for any $\e > 0$ there is 
a constant $C = C(\e)$ such that $\N\left(C(\R,\R^3) \smallsetminus S_C\right) \leq \e$ 
where
\begin{equation}
S_C = \{ Q \in C(\R,\R^3): | q_t| \; \leq \; C_0 (\ln (|t| + 1))^{1/(\alpha+1)} + C \}.
\label{co2}
\end{equation}
Then for $ Q \in S_C$ we estimate 
\begin{eqnarray*}
\lefteqn{
\int_{-T}^T ds \int_{T^*}^\infty \frac{dt}{(| q_s|^2 + (|s|+|t|)^2) \ln t 
(\ln\ln t)^\zeta} \;\geq\; } 
\\ &&
\int_{-T}^T ds \int_{T^*}^\infty \frac{dt}{((C_0 (\ln (|s|+1))^{1/(\alpha+1)} + C)^2 + 
(|s|+|t|)^2) \ln t (\ln\ln t)^\zeta}
\end{eqnarray*}
The right hand side of this last expression goes to infinity as $T \to \infty$. Hence, 
for any $\delta > 0$ we can find some $\bar T = \bar T(\delta,C)$ such that 
\begin{equation}
\mathbb{E}_{\pp_T^{ Q}} [F] \;\leq\; \delta, \;\;\; \forall  Q \in S_C, \;\; T > \bar T.
\label{co3}
\end{equation}
Moreover, it follows from (\ref{aaa}) and (\ref{bbb}) that there is a constant $B$ such 
that $\mathbb{E}_{\pp_T^{ Q}}[F] \;\leq\; B$ for all $ Q$ and $T$. Hence, by combining 
(\ref{co1}), (\ref{co2}) and (\ref{co3}), we have
\begin{equation}
\mathbb{E}_{\pp_T}[F] = \mathbb{E}_{\N_T}\left[\mathbb{E}_{\pp_T^{ Q}}[F]\right] \to 0, 
\;\;\; \mbox{as} \;\;\; T \to \infty
\end{equation}
Consider now the function
\begin{equation}
F_{\bar C} =
\left\{
\begin{array}{ll}
\vspace{0.2cm}
F(\xi), & \mbox{if $F(\xi) \leq \bar C$} 
\nonumber \\
\bar C, & \mbox{otherwise}
\end{array} \right.
\end{equation}
Clearly, for $T \to \infty$, $\mathbb{E}_{\pp_T}[F_{\bar C}] \to 0$. On the other hand, 
$\mathbb{E}_{\pp_T} [F_{\bar C}] \to \mathbb{E}_{\pp}[F_{\bar C}]$ with $T \to \infty$, 
for any $\bar C$. This implies then that $\mathbb{E}_{\pp}[F_{\bar C}] = 0$ for any 
$\bar C$ and hence $\mathbb{E}_{\pp}[F] = 0$. This completes the proof Lemma \ref{ldiv}. 
\end{pf}

\vspace{0.3cm}
\noindent
{\it Proof of Theorem \ref{tdiv}:} It follows straightforward by combining Lemma \ref{ldiv} 
with Theorem \ref{tir}. $\Box$

The smallness of the coupling constant was used in (\ref{co2}) referring to the typical 
fluctuations in the Gibbs measure derived in \cite{LM}. If we only want to show that $H$ 
has no ground state in $\h^0$, here is an alternative proof avoiding all restriction to 
small coupling. 
\begin{theorem}
Suppose $\rh \geq 0$ and $V$ is of class (P2). Then for $d=3$ $H$ has no ground state in 
$\h^0$.
\end{theorem} 
\begin{pf}
By (\ref{kkee}), (\ref{kke}) and (\ref{ekke}) we have that
\begin{equation}
(1,\Psi_T) = \frac{\mathbb{E}_{\pp^0}[Z^+_T(q,\xi)]}{\sqrt {Z_T}} =
             \frac{\mathbb{E}_{\pp^0}[Z^-_T(q,\xi)]}{\sqrt {Z_T}} 
\label{eekk}
\end{equation}
Furthermore
\begin{equation}
\mathbb{E}_{\pp^0}[Z^+_T(q,\xi)] = 
\mathbb{E}_{\n^0} \left[ \mathbb{E}_{\N^0_+}\left[\mathbb{E}_{\G}[e^{-\int_0^T (\xi_t * \rh)
(q_t) dt}] | q_0 = q \right] \right]
\end{equation}
where $\N^0_+$ is the distribution of paths in the forward direction $Q^+ = \{q_t: t \geq 0\}$.
A similar expression holds for $\mathbb{E}_{\pp^0}[Z^+_T(q,\xi)]$ taken with respect to $\N^0_-$
for paths in the backward direction $Q^- = \{q_t: t \leq 0\}$. Using the Markov property of $\N^0$
and Schwartz inequality, we obtain by (\ref{eekk}) 
\begin{equation}
(1,\Psi_T)^2 \;\leq\; \frac{1}{Z_T} \mathbb{E}_{\N^0} \left[ \mathbb{E}_{\G} [e^{-\int_{-T}^0
(\xi_t * \rh)(q_t) dt}] \; \mathbb{E}_{\G}[ e^{-\int_0^T (\xi_t * \rh)(q_t) dt}] \right].
\end{equation}
Moreover, since
\begin{equation}
\mathbb{E}_{\G}\left[ e^{-\int_0^T (\xi_t * \rh)(q_t) dt}\right] = e^{-\int_0^T\int_0^T 
W(q_t-q_s,t-s) dtds},
\end{equation}
and by a similar expression for the $[-T,0]$ interval, we get
\begin{eqnarray}
(1,\Psi_T)^2 
& \leq &
\frac{1}{Z_T} \mathbb{E}_{\N^0} \left[ e^{-\int_0^T\int_0^T W(q_t-q_s,t-s) dsdt}
e^{-\int_{-T}^0\int_{-T}^0 W(q_t-q_s,t-s) dsdt} \right] \nonumber \\
& = &
\frac{1}{Z_T} \mathbb{E}_{\N^0} \left[ e^{-\int_{-T}^T\int_{-T}^T W(q_t-q_s,t-s)dsdt +
\int\int_{\Delta_T} W(q_t-q_s,t-s) dtds} \right] \nonumber \\
& = &
\frac{1}{Z_T} \mathbb{E}_{\N} \left[ e^{\int\int_{\Delta_T} W(q_t-q_s,t-s)dsdt} \right]
\end{eqnarray} 
Here $\Delta_T = (-T,0) \times (0,T) \;\cup\; (0,T) \times (-T,0)$. 

In position space the interaction potential reads
\begin{equation}
W(q,t) = -\frac{\pi}{2} \int_{\R^3} dx \int_{\R^3} dy \frac{\rh(x)\rh(y)}{(q+x-y)^2 + t^2} 
\; < \; 0.
\end{equation}
To show that the left hand side of (\ref{eekk}) converges to zero, we first restrict to the set
\begin{equation}
A_T := \{ |q_t| \;\leq\; T^\lambda, \; \forall |t| \leq T \}
\end{equation}
with some $\lambda < 1$. By using the estimate
\begin{equation}
(q_t-q_s + x-y)^2 + (t-s)^2 \;\leq\; 8T^{2\lambda} + 2(x-y)^2 + (t-s)^2,
\end{equation}
we obtain 
\begin{eqnarray}
\lefteqn{
\int\int_{\Delta_T} dt ds \int_{\R^3} dx \int_{\R^3} dy 
\frac{\rh(x)\rh(y)}{(q_t - q_s +x-y)^2 + (t-s)^2} }  \nonumber \\ &&
\geq 2 \int_0^T dt \int_0^T ds \int_{\R^3} dx \int_{\R^3} dy \rh(x) \rh(y) 
\frac{1}{8T^{2\lambda} + 2(x-y)^2 + (t+s)^2} \nonumber \\ &&
= \int_{\R^3} dx \int_{\R^3} dy \rh(x) \rh(y) \log \frac{8T^{2\lambda} + 
2(x-y)^2 + T^2}{8T^{2\lambda} + 2(x-y)^2}.
\label{rhss}
\end{eqnarray}
The right hand side in (\ref{rhss}) goes to infinity as $T\to\infty$, since $\lambda < 1$.
Hence we have shown that
\begin{equation}
\lim_{T\to\infty} \mathbb{E}_{\N_T} \left[ \exp\left(\int\int_{\Delta_T} 
W(q_t-q_s,t-s) dsdt\right) 1_{A_T}\right] = 0.
\end{equation}

We still need to prove that the limit is zero also on the complement of $A_T$. We use again
that $W(q,t) < 0$ and estimate
\begin{equation}
\mathbb{E}_{\N_T} \left[ \exp\left(\int\int_{\Delta_T} W(q_t-q_s,t-s) dsdt\right) (1-1_{A_T})
\right] \;\leq\; \mathbb{E}_{\N_T} \left[(1-1_{A_T}) \right].
\end{equation}
Writing out the expectation with respect to $\N_T$ and by using Schwartz inequality, we obtain
\begin{eqnarray} 
\mathbb{E}_{\N_T} \left[(1-1_{A_T}) \right] 
& = &
\frac{\mathbb{E}_{\N^0} \left[e^{-\int_{-T}^T\int_{-T}^T W(q_t-q_s,t-s)dtds} (1-1_{A_T}) \right]} 
{\mathbb{E}_{\N^0} \left[e^{-\int_{-T}^T\int_{-T}^T W(q_t-q_s,t-s)dtds}\right]} \\ 
& \leq &
\frac{\left(\mathbb{E}_{\N^0} \left[e^{-2\int_{-T}^T\int_{-T}^T W(q_t-q_s,t-s)dtds}\right]\right)^{1/2}} 
{\mathbb{E}_{\N^0} \left[e^{-\int_{-T}^T\int_{-T}^T W(q_t-q_s,t-s)dtds}\right]} \left(\mathbb{E}_{\N^0} 
\left[(1-1_{A_T}) \right]\right)^{1/2} \nonumber
\end{eqnarray}
Since 
\begin{equation}
0 \;<\; - 2 \int_{-T}^T\int_{-T}^T W(q_t-q_s,|t-s|) dsdt \;<\; T \int_{\R^3}\frac{|\hat \rh(k)|^2}
{|k|^2} dk
\end{equation}
and $\int_{\R^3} dk |\hat\rh|^2/|k|^2 < \infty$, the first factor above is bounded by $\exp(aT)$ 
with some $a>0$. To conclude we have to show that this exponential growth is compensated by the 
second factor. The estimate 
\begin{equation}
\mathbb{E}_{\N^0} \left[(1-1_{A_T}) \right] = \N^0 \left( \{q_t: \; \sup_{t \in [-T,T]} |q_t| 
\;\geq\; T^\lambda \} \right) \;\leq\; f(T) e^{-c T^{\lambda(\alpha+1)}}
\label{var}
\end{equation}
is a slight modification of Lemma 5.2 in \cite{LM} based on Varadhan's lemma in conjunction with 
the bound $\exp(-c|x|^{\alpha + 1})$ on the decay of $\psi_0$. Here $c>0$, $f(T)$ is a polynomially 
growing correction, and $\alpha > 1$ is the exponent appearing in (\ref{vasy}). By choosing $1/2 < 
\lambda < 1$, the right hand side of (\ref{var}) goes to zero with $T \to \infty$.  This completes the 
proof. 
\end{pf}

We conclude this section by showing that even though for $d=3$ there is no ground state 
in Fock space, the particle remains well localized in the physical ground state. (Further 
properties of the ground state are discussed in \cite{BHLS}.) Denote
\begin{equation}
\tilde\Lambda_T (q) = \frac{d\n_T}{d\n^0} (q) 
\label{rhq}
\end{equation}
and set
\begin{equation}
\Lambda_T(q) = \tilde \Lambda_T (q) \psi_0(q)^2,
\end{equation}
as the particle density with respect to Lebesgue measure in the approximate ground state
($\psi_0$ is the ground state of the Schr\"odinger operator (\ref{schr})).
\begin{theorem}
Suppose the potential $V$ satisfies (P1). Then for sufficiently small couplings $0 < e \leq e^*$ there 
exist some constants $c_2 > c_1 >0$ such that the limit $\Lambda(q) = \lim_{T\to\infty} \Lambda_T (q)$ 
exists and 
\begin{equation}
c_1 \psi_0^2(q) \;\leq\; \Lambda (q) \;\leq\; c_2 \psi_0^2(q)
\label{crc}
\end{equation}
for every $q \in \R^d$. 
\end{theorem}
\begin{pf}
The pointwise existence of $\lim_{T\to\infty} d\n_T/d\n^0 = d\n/d\n^0$ (see (3) in Theorem 
\ref{t2}) and the equality 
\begin{equation}
\tilde \Lambda(q) := \lim_{T\to\infty} \tilde\Lambda_T(q) = \frac{d\n}{d\n^0}(q)
\label{fhw}
\end{equation}
combined with estimate (\ref{abco}) gives the result.
\end{pf}
\vspace{0.2cm}
\noindent

\section{The infrared convergent case: $d \geq 4$}

\begin{theorem}
Suppose $d \geq 4$ and that $V$ satisfies (P1). Then for sufficiently weak couplings $0 < e < e^*$ 
$H$ has a unique strictly positive ground state $\Psi$ in $\h^0$ at eigenvalue $E_0$ (and hence 
$H_{\mathrm{\tiny{N}}}$ has a unique ground state in $L^2(\R^d,dq) \otimes {\cal F}_{\mathrm{sym}}$). 
Moreover, there exists a unitary map $\Ga: \h \to \h^0$ such that $\Ga^{-1} (H-E_0) \Ga = 
H_{\mathrm{\tiny{euc}}}$.
\label{tcon}
\end{theorem}
\begin{pf}
Using the representation for $d\pp^Q_T/d\g$ given by (\ref{den}) note that there is a 
constant $C > 0$ such that
\begin{equation}
|\hat m_T(k;Q)| \;\leq\; C \frac{|\hat \rh(k)|}{|k|} 
\label{uniqt}
\end{equation}
uniformly in $Q$ and $T$, and hence it follows that the functional $\int_{\R^d} \xi(x) 
m_T(x;Q) dx$ is well defined for $\g$-almost all $\xi \in \B_{D}$. Similarly, we have
\begin{equation}
\int_{\R^d} |\hat g_T^0(k;Q)|^2 |k| dk \; \leq \; C \int_{\R^d} 
\frac{|\hat \rh(k)|^2}{|k|^3} dk  < \infty
\end{equation}
uniformly in $Q$ and $T$. Also, the density
\begin{equation}
\frac{d\pp^{Q}}{d\g} = \exp \left(\int_{\R^d} \xi(x) m(x;Q) dx - \frac{1}{4} \int_{\R^d} 
|\hat g^0(k;Q)|^2 |k|dk \right)
\label{denex}
\end{equation}
exists, where
\begin{eqnarray*}
&& m(x;Q) = \int_{\R^d} \hat m(k;Q) e^{ik\cdot x} dk,   \\
&& \hat m(k;Q) = |k| \hat g^0(k;Q). 
\end{eqnarray*}
We write now the Radon-Nikodym derivative $d\pp_T/d\pp^0$ in another way, that is
\begin{equation}
\frac{d\pp_T}{d\pp^0} (q,\xi) = 
\mathbb{E}_{\N_T} \left [\frac{d\pp^Q_T}{d\g}(\xi)|q_0 = q \right] \frac{d\n_T}{d\n^0}(q)
\label{fubi}
\end{equation}
for every $0 < T < \infty$. By Fubini's theorem for any $q \in \R^d$ and $0 < T \leq \infty$ 
there is a set $\Omega = \Omega(q,T) \subset \B_{D}$, $\pp^0(\Omega) = 1$, such that for 
$\xi \in \Omega$ the conditional expectation $\mathbb{E}_{\N_T}[d\pp^Q_T/d\pp^0(\xi)|q_0 = q]$ 
exists and hence (\ref{fubi}) is well defined. 

Next, from (\ref{fubi}) we see
\begin{equation}
\mathbb{E}_{\pp^0}\left[\left(\frac{d\pp_T}{d\pp^0}(q,\xi) \right)^{1/2}\right] = 
\int \left( \mathbb{E}_{\N_T} \left [\frac{d\pp^Q_T}{d\g}(\xi)|q_0 = q \right] \right)^{1/2} 
\left(\frac{d\n_T}{d\n^0}(q)\right)^{1/2} d\g d\n^0.
\end{equation}
We apply Jensen's inequality and use (\ref{den}) to obtain
\begin{eqnarray*}
\lefteqn{
\left(\mathbb{E}_{\N_T}\left [\frac{d\pp^Q_T}{d\pp^0}(\xi)|q_0 = q \right]\right)^{1/2} \;\geq\; } 
\\ &&
\exp\left(\frac{1}{2} \int_{\R^d} \xi(x) \mathbb{E}_{\N_T} \left[m_T(x;Q)|q_0 = q \right] dx - 
    \frac{1}{8} \int_{\R^d} \mathbb{E}_{\N_T} \left[|\hat g_T^0(k;Q)|^2 \;| \;q_0 = q \right]
|k|dk\right).
\end{eqnarray*}
By using once again Jensen's inequality and the fact that the field has zero mean under $\g$, we 
find
\begin{eqnarray}
\lefteqn{
\mathbb{E}_{\pp^0} \left[ \left(\frac{d\pp_T}{d\pp^0}(q,\xi)\right)^{1/2} \right] \;\geq\; } 
\\ &&
\int \exp \left(-\frac{1}{8} \int_{\R^3} \mathbb{E}_{\N_T} \left[|\hat g_T^0 (k;Q)|^2|q_0 = q 
\right] 
|k| dk \right) \left(\frac{d\n_T}{d\n^0}(q)\right)^{1/2} d\n^0(q). \nonumber
\end{eqnarray}
Estimate (\ref{not-s}) gives then
\begin{equation}
\mathbb{E}_{\N_T}\left[|\hat g_T^0(k;Q)|^2|q_0 = q \right] \;\leq\; C \frac{|\hat\rh(k)|^2}{|k|^4}.
\label{posi}
\end{equation}
Since by (\ref{abco}) $d\n_T/d\n^0$ is bounded from below uniformly in $T$ and $q$, and by using
Lemma \ref{lemm} and the Remark following it we arrive at
\begin{equation}
\mathbb{E}_{\pp^0} \left[ \left(\frac{d\pp_T}{d\pp^0}(q,\xi)\right)^{1/2} \right] \;\geq\; C_1 
\; \exp \left(-C_2 \int_{\R^d} \frac{|\hat\rh(k)|^2}{|k|^3} dk \right) \;\geq\; C_3 \;>\;  0.
\label{ccc}
\end{equation}
By (\ref{ccc}) it then follows that the left hand side above cannot converge to zero and, as it 
follows from Lemmas \ref{lem} and \ref{lemm}, there exists a function $\Psi \in \h^0$ such that 
\begin{equation}
\lim_{T\to\infty} \frac{d\pp_T}{d\pp^0} = \Psi^2 = \frac{d\pp}{d\pp^0}.
\label{qqqq}
\end{equation}
Furthermore we use formulas (\ref{fubi}) and (\ref{denex}) and apply once more Jensen's inequality 
to obtain
\begin{eqnarray*}
\lefteqn{
\frac{d\pp}{d\pp^0}(q,\xi) \geq } \\ &&
\exp\left( \int_{\R^d} \xi(x) \mathbb{E}_{\N}[m(x;Q)|q_0=q] dx - \frac{1}{8} \int_{\R^d}
\mathbb{E}_{\N}[|\hat g^0(k;Q)|^2|q_0=q] |k| dk \right) \frac{d\n}{d\n^0}(q). 
\end{eqnarray*}
Since the function 
\begin{equation}
\tilde m(x;q) = \mathbb{E}_{\N} [m(x;Q)|q_0=q]
\end{equation}
satisfies the estimate (\ref{uniqt}) it is easy to show that $\exp(\xi(x) \tilde m(x;q)) > 0$ 
for $\pp^0$-almost all $(q,\xi) \in \R^d \times \B_D$. By (\ref{posi}) and the positivity of 
$d\n/d\n^0$ we get that $d\pp/d\pp^0(q,\xi) > 0$ for $\pp^0$-almost all $(q,\xi)$. 
By Lemma \ref{lemm} (i) we have then that the ground state $\Psi$ exists. 

We conclude by showing that there is a unitary map $\Ga: \h \to \h^0$, $ F \mapsto F \Psi$ 
transforming $H - E_0$ into $H_{\mbox{\tiny{euc}}}$. Indeed, by using (\ref{fkh}) and (\ref{euc}) 
\begin{eqnarray*}
(F, T_t G)_{\h} 
& = &
\mathbb{E}_{\p} [F(q_0,\xi_0) G(q_t,\xi_t)] \\ 
& = &
\lim_{T\to\infty} \frac{1}{Z_{2T+t}} \mathbb{E}_{\p^0} [e^{-\int_{-T}^{T+t} (\xi_s * \rh)(q_s) ds} 
F(q_0,\xi_0) G(q_t,\xi_t) ] \\
& = &
\lim_{T\to\infty}\frac{(e^{-TH}1, F e^{-tH}Ge^{-TH}1)_{\h^0}}{(1,e^{-(2T+t)H}1)_{\h^0}} \\
& = &
\frac{\int \Psi(q_0,\xi_0) F(q_0,\xi_0)[ e^{-tH}G \Psi](q_0,\xi_0) d\pp^0}{e^{-tE_0}} \\
& = &
(\Gamma F, e^{-t(H-E_0)}\Gamma G)_{\h^0} \\
& = &
(F,\Gamma^{-1} e^{-t(H-E_0)} \Gamma G)_{\h}.
\end{eqnarray*}
This completes the proof of the theorem. 
\end{pf}

The proof above requires a uniform lower bound on $d\n_T/d\n^0$, which is at present available only
under the strong assumptions (P1) and $0 < e \leq e^*$. In \cite{Spo2} a uniform lower bound on 
$(\Psi_T, e^{-\tau H_{\rm{\tiny{p}}}} \otimes P_{\Omega} \Psi_T)$ has been obtained by using, as here,
a functional integral representation ($P_\Omega$ is the projection onto the Fock vacuum). In our 
context this technique proves the existence of the ground state of $H_{\rm{\tiny{N}}}$ for $d \geq 4$ 
with assumption (P2) on the potential and with no restriction on $e$. 

\section{Absence of gap in the spectrum of $H_{\mbox{\tiny{euc}}}$}
Consider the subspace $\hat\h \subset \h$ of functions having the property that 
$\mathbb{E}_{\pp}[f] = 0$. Obviously, this subspace is invariant under the semigroup generated 
by $H_{\mathrm{\tiny{euc}}}$. We denote the restriction of this operator to $\hat\h$ by 
$\hat H_{\mbox{\tiny{euc}}}$. 
\begin{theorem}
For any $d \geq 3$ we have $\inf \Spec \hat H_{\mbox{\tiny{euc}}} = 0$. 
\end{theorem}
\begin{pf}
Consider the functional on $\h$ 
\begin{equation}
F_h (\xi) = \int_{\R^d} \xi(x) h(x) dx := \xi(h), \;\;\; h \in {\cal S}(\R^d).
\end{equation}
Clearly, $\xi(h) - \mathbb{E}_{\pp}(\xi(h)) := \tilde F(h)$ is an element of $\hat\h$, and 
\begin{equation}
(e^{-\hat H_{\mbox{\tiny{euc}}}t} \tilde F_h, \tilde F_h)_{\h}  = \int_0^\infty e^{-t\lambda} 
d\sigma_h(\lambda) := G_h(t)
\end{equation}
with
\begin{equation}
d\sigma_h(\lambda) = d(\tilde F_h,\hat P_\lambda \tilde F_h) 
\end{equation}
and $\{\hat P_\lambda\}$ the spectral family of $\hat H_{\mbox{\tiny{euc}}}$. Lemma \ref{lcor} 
below implies that $\inf \supp d\sigma_h(\lambda) = 0$, and this on its turn implies 
$\inf \sup \hat P_\lambda = 0$. 
\end{pf}

\begin{lemma}
The following asymptotics holds for $t \to \infty$,
\begin{equation}
G_h(t) = \frac{C |\hat h(0)|^2}{t^{d-1} + 1} + o \left(\frac{1}{t^{d-1} + 1} \right)
\end{equation}
\label{lcor}
with some constant $C > 0$. 
\end{lemma}
\begin{pf}
\begin{eqnarray*}
(e^{-\hat H_{\mbox{\tiny{euc}}}t} \tilde F_h, \tilde F_h)_{\h}
& = &
\cov_{\p} (\xi_t(h), \xi_0(h)) \\
& = &
\mathbb{E}_{\N}\left[\mathbb{E}_{\p^{Q}}\left[(\xi_t(h) - \mathbb{E}_{\p^{Q}}[\xi_0(h)])
(\xi_0(h) - \mathbb{E}_{Q}[\xi_0(h)])\right]\right] \\
& + &
\mathbb{E}_{\N}\left[(\mathbb{E}_{\p^{Q}}[\xi_t(h)] - \mathbb{E}_{\p}[\xi_t(h)])
( \mathbb{E}_{\pp^{ Q}}[\xi_0(h)] - \mathbb{E}_{\p}[\xi_0(h)])\right] 
\end{eqnarray*}
Here we used the identity 
\begin{eqnarray}
\mathbb{E}_{\mu}\left[ (F - \mathbb{E}_{\mu}[F])(G - \mathbb{E}_{\mu}[G]) \right] 
& = &
\mathbb{E}_{\mu}\left[ (F - \mathbb{E}_{\mu}[F|{\cal F}])(G - \mathbb{E}_{\mu}[G|{\cal F}]) 
\right] 
\label{bibi} \\
& + &
\mathbb{E}_{\mu}\left[ (\mathbb{E}_{\mu}\left[F|{\cal F}\right] - \mathbb{E}_{\mu}[F])
(\mathbb{E}_{\mu} \left[G|{\cal F}\right] - \mathbb{E}_{\mu}[G]) \right] \nonumber
\end{eqnarray}
valid for an arbitrary probability measure $\mu$, where $\mathbb{E}_\mu(\;\cdot\;|{\cal F})$ 
is a conditional average with respect to some $\sigma$-field $\cal F$, and $F$ and $G$ are 
given random variables. Since for any fixed trajectory $Q$ the conditional covariance of 
$\xi_t(x)$ is the same as before, we get
\begin{equation}
\mathbb{E}_{\p^{ Q}} \left[(\xi_t(h) - \mathbb{E}_{\p^{Q}} [\xi_t(h)])(\xi_0(h) - 
\mathbb{E}_{\p^{Q}} [\xi_0(h)])\right] = \int_{\R^d} \frac{e^{-|k| |t|}}{|k|} |\hat h(k)|^2 dk. 
\label{we}
\end{equation}
Then by choosing a sufficiently rapidly decreasing smooth function $h$ with $\hat h(0) = 
\int h(x) dx \neq 0$, we see that the integral (\ref{we}) behaves asymptotically like 
$\mbox{const}|\hat h(0)|^2/t^{d-1} + o(1/t^{d-1})$. 

We turn next to the second term in (\ref{bibi}). We have
\begin{equation}
\mathbb{E}_{\p^{ Q}} [\xi_t(h)] = -\frac{1}{4} \int_{-\infty}^\infty \int_{\R^d} \hat h(k) 
\frac{\hat \rh(k)}{|k|} e^{-ik\cdot q_\tau} e^{-|k| |t-\tau|} d\tau dk
\end{equation}
and by stationarity
\begin{equation}
\mathbb{E}_{\p}[\xi_t(h)] = -\frac{1}{4} \int_{\R^d} dk \int_{-\infty}^\infty \hat h(k) 
\mathbb{E}_{\N}[e^{-ik\cdot q}] \frac{\hat\rh(k)}{|k|} e^{-|k| |t-\tau|} d\tau.
\end{equation}
Hence
\begin{equation}
\mathbb{E}_{\p^{ Q}} [\xi_t(h)] - \mathbb{E}_{\p} [\xi_t(h)] = -\frac{1}{4} \int_{-\infty}^\infty 
d\tau \int_{\R^d} \hat h(k) \frac{\hat \rh(k)}{|k|} e^{-|k| |t-\tau|} (e^{-ik\cdot q_\tau} -
\mathbb{E}_{\N}[e^{-ik\cdot q}])dk. 
\end{equation}
Furthermore
\begin{eqnarray*}
\lefteqn{
\left(\mathbb{E}_{\p^{Q}} [\xi_t(h)] - \mathbb{E}_{\p}[\xi_t(h)]\right) 
\left(\mathbb{E}_{\p^{Q}} [\xi_0(h)] - \mathbb{E}_{\p}[\xi_0(h)]\right) = } \\ &&
\frac{1}{16} \int_{-\infty}^\infty \int_{-\infty}^\infty d\tau_1 d\tau_2 \int_{\R^d}\int_{\R^d} 
dk_1 dk_2 \hat h(k_1) \hat h(k_2) \frac{\hat \rh(k_1) \hat \rh(k_2)}{|k_1| |k_2|} 
e^{-|k| |t-\tau_1|} e^{-|k_2| |\tau_2|} \times \\ && 
\hspace{1cm} \times 
\left(e^{-ik_1 \cdot q_{\tau_1}} - \mathbb{E}_{\N}[e^{-ik_1 \cdot q}] \right)
\left(e^{-ik_2 \cdot q_{\tau_2}} - \mathbb{E}_{\N}[e^{-ik_2 \cdot q}] \right)
\end{eqnarray*}
Making use of (\ref{deca}) on the decay of correlations for $\N$ we get that
\begin{eqnarray}
\lefteqn{
\mathbb{E}_{\N}\left[(\mathbb{E}_{\p^{ Q}}[\xi_t(h)] - \mathbb{E}_{\p} [\xi_t(h)])
     (\mathbb{E}_{\p^{Q}} [\xi_0(h)] - \mathbb{E}_{\p} [\xi_0(h)]) \right] \label{deen}} \\ &&
\leq \; \bar C \int_{-\infty}^\infty \int_{-\infty}^\infty d\tau_1 d\tau_2 
\int_{\R^d} \int_{\R^d} \hat h(k_1) \hat h(k_2) \frac{\hat\rh(k_1) \hat\rh(k_2)}{|k_1| |k_2|} 
\frac{e^{-|k_1| |t-\tau_1| - |k_2| |\tau_2|}}{|\tau_1 - \tau_2|^\gamma + 1} dk_1 dk_2 
\nonumber \\ &&
\leq \; \bar C (\sup |\hat h|)^2 \int_{-\infty}^\infty \int_{-\infty}^\infty \frac{d\tau_1 d\tau_2}
{(|t-\tau_1|^{d-1} + 1) (|\tau_2|^{d-1} + 1)(|\tau_1-\tau_2|^\gamma + 1)} \nonumber
\end{eqnarray}
with some $\bar C >0$. Denote 
\begin{equation}
L_1(t) = \frac{1}{|t|^{d-1} + 1}, \;\;\; L_2(t) = \frac{1}{|t|^\gamma + 1}
\end{equation}
The right hand side of (\ref{deen}) above is the double convolution $L_1 * L_1 * L_2$. Then we 
get by Fourier transformation
\begin{equation}
(\widehat{L_1 * L_1 * L_2})(y) = \hat L_1(y)^2 \hat L_2(y), \;\;\; y \in \R.
\end{equation}
By general results on the asymptotics of Fourier transforms \cite{AGV} we have for small $y$ 
\begin{eqnarray*}
&& \hat L_1(y) \sim |y|^{d-2} \\
&& \hat L_2(y) \sim |y|^{\gamma -1}
\end{eqnarray*}
Thus $\hat L_1(y)^2 \hat L_2(y) \sim |y|^{2d + \gamma -5}$. The asymptotics of the integral at the
right hand side of (\ref{deen}) becomes then $1/t^{2d + \gamma - 4}$. Note that for $d \geq 3$ and 
$\gamma > 0$, $2d + \gamma - 4 > d-1$, and thus the second term in (\ref{bibi}) is $o(1/t^{d-1})$. 
This completes the proof.
\end{pf}

\vspace{0.5cm}
{\bf Acknowledgments:} J.L. is supported by Deutsche Forschungs\-gemeinschaft within 
Schwerpunktprogramm ``Interagierende stochastische Systeme von hoher Komplexit\"at''. R.A.M. thanks 
Zentrum Mathematik of Technische Universit\"at M\"unchen for warm hospitality and financial support. 
He also thanks the Russian Fundamental Research Foundation (grants 99-01-00284 and 00-01-00271) for 
financial support.

\end{document}